\documentclass[submission,copyright,creativecommons]{eptcs}

\providecommand{\event}{WLP 2015} 
\usepackage{breakurl}             
\usepackage{underscore}           

\usepackage{epsfig}
\usepackage{epsfig}
\usepackage{listings}

\usepackage{graphicx,amssymb,lineno,color}

\makeatletter
\def\elsartstyle{%
    \def\normalsize{\@setfontsize\normalsize\@xiipt{14.5}}
    \def\small{\@setfontsize\small\@xipt{13.6}}
    \let\footnotesize=\small
    \def\large{\@setfontsize\large\@xivpt{18}}
    \def\Large{\@setfontsize\Large\@xviipt{22}}
    \skip\@mpfootins = 18\p@ \@plus 2\p@
    \normalsize
}
\makeatother

\usepackage{times}
\usepackage{helvet}
\usepackage{courier}
\usepackage{amsmath}

\usepackage[T1]{fontenc}
\usepackage[latin1]{inputenc}
\fontfamily{ptm}\selectfont
\usepackage{url}
\usepackage{xspace}
\usepackage{calc}
\usepackage{verbatim} 
\usepackage{paralist}

\newcommand{\swi}{\textsc{Swi}\xspace}
\newcommand{\prolog}{\textsc{Prolog}\xspace}
\newcommand{\datalog}{\textsc{Datalog}\xspace}
\newcommand{\datalogs}{\textsc{Datalog}$^\star$\xspace}

\newcommand{\codett}[1]{\texttt{#1}}

\newcommand{\owl}{\textsc{Owl}\xspace}
\newcommand{\swrl}{\textsc{Swrl}\xspace}
\newcommand{\rdf}{\textsc{Rdf}\xspace}
\newcommand{\ruleml}{\textsc{RuleMl}\xspace}
\newcommand{\odbc}{\textsc{Odbc}\xspace}
\newcommand{\ddbase}{\textsc{DDbase}\xspace}
\newcommand{\ddk}{\textsc{Ddk}\xspace}
\newcommand{\RuleML}{\textsc{RuleML}\xspace}
\newcommand{\XML}{\textsc{Xml}\xspace}
\newcommand{\java}{{\sc Java}\xspace}

\newcommand{\Rulearrow}{ \leftarrow }
\newcommand{\Rule}[2]{%
   \mathit{#1} \leftarrow #2 }

  \newcommand{\ts}{\hspace*{ 0.25mm}}

\newcommand{\tuple}[1]{\langle\ts\ts #1 \ts\ts\rangle}

\newcommand{\eat}[1]{}

\newenvironment{mquote}{%
   \begin{quote} \( }{ \)
   \end{quote}}
\newenvironment{dqnarray*}{%
   \ddbbook{\vspace*{2mm}}
   \begin{quote}
       \( \begin{array}{lcl}}{\end{array} \)
   \end{quote} }
\newenvironment{mqnarray*}{%
   \vspace*{2mm}
   \begin{quote}
      \( \begin{array}{l}}{\end{array} \)
   \end{quote} }

  \newcommand{\NamedRule}[3]{%
     \mathit{#1} \ \mbox{\small $=$} \ #2 \leftarrow #3 }
  \newcommand{\dand}{\wedge}
  \newcommand{\defneg}{{\sf not}\ }
  \newcommand{\defnegs}{\mathit{not}}

  \newcommand{\mgu}{\textsc{Mgu}\xspace}

  \newcommand{\bi}{\begin{itemize}}
  \newcommand{\ei}{\end{itemize}}

  \newcommand{\fnquery}{\textsc{FnQ}uery\xspace}
  \newcommand{\sql}{\textsc{Sql}\xspace}
  \newcommand{\xml}{\textsc{Xml}\xspace}

\definecolor{comment}{rgb}{0.2, 0.6, 0.4}
\definecolor{string}{rgb}{0, 0, 0}
\definecolor{keyword}{rgb}{0, 0, 0}
\definecolor{background}{rgb}{1, 1, 1}
\definecolor{lightgrey}{rgb}{0.95, 0.95, 0.98}

\title{%
   Knowledge Engineering for
   Hybrid Deductive Databases%
}

\author{%
   Dietmar Seipel
   \institute{%
      University of W\"urzburg,
      Institute for Computer Science, Germany \\[1mm]
      \email{seipel@informatik.uni-wuerzburg.de}
   }
}

\def\titlerunning{%
   Knowledge Engineering for
   Hybrid Deductive Databases%
}
\def\authorrunning{Dietmar Seipel}

\begin{document}

\maketitle

\begin{abstract}
Modern knowledge base systems frequently need to combine
a collection of databases in different formats:
e.g., relational databases, \xml databases,
rule bases, ontologies, etc..
In the deductive database system \ddbase, we can manage
these different formats of knowledge and reason about them.
Even the file systems on different computers can be part
of the knowledge base.
Often, it is necessary to handle different versions
of a knowledge base.
E.g., we might want to find out common parts or differences
of two versions of a relational database.

We will examine the use of abstractions of rule bases
by predicate dependency and rule predicate graphs.
Also the proof trees of derived atoms can help to
compare different versions of a rule base.
Moreover, it might be possible to have
derivations joining rules with other formalisms of
knowledge representation.

Ontologies have shown their benefits in many applications
of intelligent systems,
and there have been many proposals for rule languages
compatible with the semantic web stack, e.g., \swrl,
the semantic web rule language.
Recently, ontologies are used in hybrid systems for
specifying the provenance of the different components.
\end{abstract}

\subsubsection*{Keywords.}
   hybrid knowledge bases, deductive databases,
   rules, ontologies

\section{Introduction}
\label{sec:Introduction}

Relational databases and deductive databases with rule bases
have been prominent formalisms of knowledge representation
for a long time \cite{CGT:90}.
A strong research interest in deductive database technology
and its applications has reemerged in the recent years,
leading to what has been called a resurgence
\cite{Hellerstein:10} or even a renaissance
\cite{Abiteboul:12} for \datalog.
This revival is propelled by
important new applications areas,
such as distributed computations and big--data applications,
the success of a first commercial \datalog System, and
progress in semantics extensions to support non--monotonic
constructs such as default negation and aggregates --
a theoretical thread that had actually continued through
the years~\cite{MiZaSei:14}.

In the last years, the use of ontologies has shown its benefits
in many applications of intelligent systems.
There have been many proposals for rule languages compatible with the
semantic web stack, e.g., the definition of \swrl (semantic web rule language)
originating from \RuleML and similar approaches~\cite{HPBT:05}.
It is well agreed that the combination of ontologies with rule--based knowledge
is essential for many interesting semantic web tasks, e.g., the realization of
semantic web agents and services.
\swrl allows for the combination of a high--level abstract syntax for Horn--like
rules with \owl, and a model theoretic semantics is given for the combination
of \owl with \swrl rules.
An \XML syntax derived from \RuleML allows for a syntactical compatibility with
\owl.
However, with the increased expressiveness of such ontologies new demands for
the development and for maintenance guidelines arise.

Thus, approaches for evaluating and maintaining \emph{hybrid}
knowledge bases need to be extended and revised to work with
different kinds of knowledge including rules,
relational databases, \xml documents, and ontologies.
In this paper, we are interested in various types of knowledge bases
including
   a collection of relational databases or ontologies,
   or the hierarchy in a file system.
Here, the interaction between the knowledge bases can be
very important.
E.g., there could be a call from a logical rule to a bayesian network.
%
%
The abstraction of knowledge bases is related to
the schemas of relational and \xml databases,
the predicate dependency and rule predicate graphs of
deductive databases and rule--based systems.
Obviously, w.r.t.\ versioning, there is a relationship to
synchronisation and diff for programs and files.
For file systems, well--known characteristics are
the size parameters, such as words, lines, characters.

\paragraph{Organization of the Paper.}

The rest of this paper is organized as follows:
Section~2 shows how rule bases can be abstracted and
visualized in \ddbase using different types
of dependency graphs.
Derivations can be explained by proof trees.
Section~3 investigates ontologies with rules
in \swrl, and it shows how the provenance of
ontologies can be modelled and reasoned about.
Hybrid knowledge bases can also be queried in
combined \prolog statements in \ddbase,
see Section~4.
The paper is concluded with some final remarks.

\section{Graphs for Rule Bases}

In a deductive database, a logic program can be abstracted
by a predicate dependency or a rule predicate graph,
and a derivation can be visualized by its proof tree,
cf.\ \cite{CGT:90}.
It should also be possible to have bottom--up and top--down
evaluation in one system.
\datalogs can evaluate logic programs with \prolog syntax
(extended \datalog programs) in a bottom--up style;
it is designed to evaluate embedded \prolog calls
in a top--down manner \cite{Sei:09}.

\subsection*{Dependency Graphs}

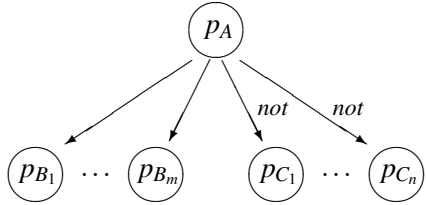
\begin{figure}[ht]
\begin{center} \unitlength 8mm
\begin{picture}(8,3.9)(3,1.2)

\put (7,4.2){\oval(0.9,0.9)}
\put (7.05,4.2){\makebox(0,0){$p_A$}}
\put (4,1.8){\oval(0.9,0.9)}
\put (4.05,1.8){\makebox(0,0){$p_{B_1}$}}
\put (5,1.8){\makebox(0,0){$\ldots$}}
\put (6,1.8){\oval(0.9,0.9)}
\put (6.05,1.8){\makebox(0,0){$p_{B_m}$}}
\put (8,1.8){\oval(0.9,0.9)}
\put (8.05,1.8){\makebox(0,0){$p_{C_1}$}}
\put (9,1.8){\makebox(0,0){$\ldots$}}
\put (10,1.8){\oval(0.9,0.9)}
\put (10.05,1.8){\makebox(0,0){$p_{C_n}$}}

\put (6.6,3.7){\vector(-3,-2){2.1}}
\put (6.9,3.7){\vector(-1,-2){0.67}}
\put (7.4,3.7){\vector(3,-2){2.1}}
\put (7.1,3.7){\vector(1,-2){0.67}}

\put (7.95,2.9){\makebox(0,0){\footnotesize $\defnegs$}}
\put (9.2,2.9){\makebox(0,0){\footnotesize $\defnegs$}}

\end{picture}
\end{center}
   \caption{Predicate Dependency Graph.}
   \label{Predicate Dependency Graph}
\end{figure}

We can define two sorts of dependency graphs
for a normal logic program $P$.
Both reflect, which predicates call other predicates in $P$.
Let $p_L$ be the predicate symbol of a literal $L$.
The \emph{predicate dependency graph} $G_P^d = \tuple{V_P^d,E_P^d}$
of $P$ is given by:
\bi
   \item the node set $V_P^d$ is the set of predicate symbols in $P$,
   \item
$E_P^d$ contains an edge $\tuple{p_A,p_{L_i}}$ for every rule
   \( \Rule{A}{L_1 \dand \ldots \dand L_m} \)
of $P$ and $1 \leq i \leq m$.
\ei
The \emph{rule predicate graph} $G_P^{rg} = \tuple{V_P^{rg},E_P^{rg}}$
of $P$ is given by:
\bi
   \item
$V_P^{rg}$ contains a node for every predicate symbol $p_L$
in $P$ and every rule $r$ in $P$,
   \item
$E_P^{rg}$ contains an edge $\tuple{p_A,r}$ and an edge
$\tuple{r,p_{L_i}}$ for every rule
   \( \NamedRule{r}{A}{L_1 \dand \ldots \dand L_m} \)
of $P$ and $1 \leq i \leq m$.
\ei
In both graphs, the edges $\tuple{p_A,p_{L_i}} \in E_P^d$ and
$\tuple{r,p_{L_i}} \in E_P^{rg}$, respectively, which come
from default negated body literals $L_i = \defneg C_i$, are
marked by ``$\defnegs$''.
We get the following dependency graphs for a normal rule
\begin{quote}
   \( \NamedRule{r}{A}{B_1 \dand \dots \dand B_m \dand
         \defneg C_1 \dand \dots \dand \defneg C_n}. \)
\end{quote}
The rule predicate graph $G_P^{rg}$ is more refined than the
predicate dependency graph $G_P^d$,
see Figure~\ref{Predicate Dependency Graph}.

\eat{
\begin{figure}[ht]
\begin{center} \unitlength 8mm
\begin{picture}(8,5.9)(3,1.2)

\put (7,6.2){\oval(0.9,0.9)}
\put (7.05,6.2){\makebox(0,0){$p_A$}}
\put (6.6,3.8){\framebox(0.8,0.8){$r$}}
\put (4,1.8){\oval(0.9,0.9)}
\put (4.05,1.8){\makebox(0,0){$p_{B_1}$}}
\put (5,1.8){\makebox(0,0){$\ldots$}}
\put (6,1.8){\oval(0.9,0.9)}
\put (6.05,1.8){\makebox(0,0){$p_{B_m}$}}
\put (8,1.8){\oval(0.9,0.9)}
\put (8.05,1.8){\makebox(0,0){$p_{C_1}$}}
\put (9,1.8){\makebox(0,0){$\ldots$}}
\put (10,1.8){\oval(0.9,0.9)}
\put (10.05,1.8){\makebox(0,0){$p_{C_n}$}}

\put (7,5.65){\vector(0,-1){0.9}}

\put (6.6,3.7){\vector(-3,-2){2.1}}
\put (6.9,3.7){\vector(-1,-2){0.67}}
\put (7.4,3.7){\vector(3,-2){2.1}}
\put (7.1,3.7){\vector(1,-2){0.67}}

\put (7.95,2.9){\makebox(0,0){\footnotesize $\defnegs$}}
\put (9.2,2.9){\makebox(0,0){\footnotesize $\defnegs$}}

\end{picture}
\end{center}
   \caption{Rule Goal Graph.}
   \label{Rule Goal Graph}
\end{figure}
}
The following definite logic programs have the same
predicate dependency graph $G_{\! P}^d$ but different
rule predicate graphs $G_{\! P_1}^{rg}$ and $G_{\! P_2}^{rg}$:
\begin{mqnarray*}
   P_1 = \{\: \Rule{p}{q_1},\, \Rule{p}{q_2}\: \}, \\
   P_2 = \{\: \Rule{p}{q_1 \dand q_2}\: \},
\end{mqnarray*}
The predicate dependency graph can be obtained from the
rule predicate graph by transitively connecting the head and
body predicate symbols and omitting the rule nodes
between.
Restricted to the predicate symbols, both graphs habe the same
paths.
They are typically used for analysing logic programs of software
packages and for refactoring and slicing.

\subsection*{Dependency Graphs with Helper Rules}

Dependency graphs can help to detect that two logic programs
are equivalent apart from helper rules:
\begin{mquote}
   \NamedRule{r_1}{A}{
      B_1 \dand \ldots \dand B_{i-1} \dand B_i \dand
      B_{i+1} \dand \ldots \dand B_n} \\
   \NamedRule{r_2}{B}{
      C_1 \dand \ldots \dand C_m}
\end{mquote}
With the most general unifier $\theta = \mgu(B_i, B)$
we can obtain the following resolvent of the two rules:
\begin{mquote}
   \NamedRule{r_3}{A\theta}{
      B_1\theta \dand \ldots \dand B_{i-1}\theta \dand
      C_1\theta \dand \ldots \dand C_m\theta \dand
      B_{i+1}\theta \dand \ldots \dand B_n\theta}
\end{mquote}
Apart from the helper predicate symbol $p_B = p_{B_i}$,
the set of predicate symbols that is reachable from $p_A$
in the predicate dependency or rule predicate graph is the same
for the two programs
   \( \{ r_1, r_2 \} \)
and
   \( \{ r_3 \}. \)
For the former, we get
   \( \Pi = \{ p_{B_1},\ldots,p_{B_n} \} \cup
      \{ p_{C_1},\ldots,p_{C_m} \}, \)
for the latter we get
   \( \Pi \setminus \{ p_B \}. \)

\subsection*{Dependency Graphs with Meta--Predicates}

In \ddbase, we use an extra node for every call
of a meta--predicate to correctly reflect a predicate.
The specific difference between normal predicates and
meta--predicates in our extension of the rule predicate graph
is that there can be several nodes labelled with
the same meta--predicate.

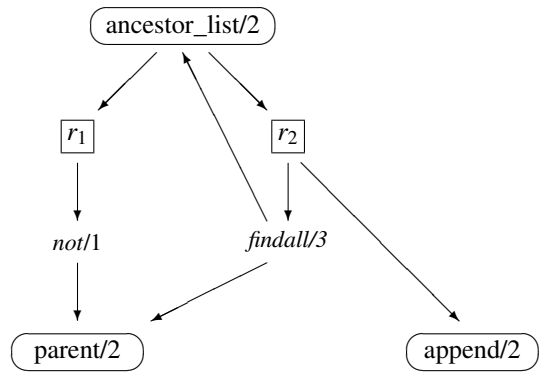
\begin{figure}[ht]
   \small
\unitlength 7mm
\begin{picture}(15,7.5)(-2,1.35)

\put(5,8.1){\oval(3.5,0.7) \makebox(0,0){ancestor\_list/2}}
\put(3,6){\makebox(0,0){\framebox(0.6,0.6){$r_1$}}}
\put(7,6){\makebox(0,0){\framebox(0.6,0.6){$r_2$}}}
\put(3,4){\makebox(0,0){\footnotesize $\defnegs$/1}}
\put(7,4){\makebox(0,0){\emph{\footnotesize findall/3}}}
\put(3,1.9){\oval(2.5,0.7) \makebox(0,0){parent/2}}
\put(10.5,1.9){\oval(2.5,0.7) \makebox(0,0){append/2}}

\put(4.5,7.6){\vector(-1,-1){1.1}}
\put(5.5,7.6){\vector(1,-1){1.1}}

\put(3,5.5){\vector(0,-1){1.1}}
\put(3,3.6){\vector(0,-1){1.1}}
\put(7,5.5){\vector(0,-1){1.1}}
\put(6.6,3.6){\vector(-2,-1){2.2}}

\put(6.6,4.4){\vector(-1,2){1.6}}

\put(7.25,5.5){\vector(1,-1){3}}

\end{picture}
   \caption{Rule Predicate Graph with Meta Predicates.}
   \label{Meta Predicates}
\end{figure}

The following example contains two such calls,
i.e.\ to the meta--predicates \codett{not/1} and \codett{findall/3}.
The predicate \codett{ancestor\_list/2} derives the list \codett{Xs}
of ancestors of a person \codett{X}.
\begin{quote}
\begin{lstlisting}{}
ancestor_list(X, [X]) :-
   not(parent(X, _)),
   !.
ancestor_list(X, Xs) :-
   findall( Ys,
      ( parent(X, Y),
        ancestor_list(Y, Ys) ),
      Yss ),
   append(Yss, Xs).
\end{lstlisting}
\end{quote}
The \prolog program is recursive, since \codett{ancestor\_list/2}
calls itself through \codett{findall/2},
see Figure~\ref{Meta Predicates}.
After that, the predicate \codett{append/2} appends the list
\codett{Yss} of derived lists to a regular list \codett{Xs}.

Obviously, the \prolog program above only terminates for
top--down evaluation on acyclic predicates \codett{parent/2}.
The nodes for the meta--prediactes are necessary to show
that \codett{ancestor\_list/2} depends on \codett{parent/2}
and \codett{ancestor\_list/2} itself.
If we would use ordinary predicate dependency or rule predicate
graphs, then we would lose this information.

\subsection*{Derivations and Proof Trees}

We are developing a deductive database system \ddbase,
which can manage hybrid rule bases and embed \prolog calls
into \datalog rules.
The underlying language \datalogs has \prolog syntax;
its mixing of bottom--up and top--down evaluation is described
in \cite{Sei:09}.
The following logic program deals with a well--known extension
of the route finding problem from deductive databases.
In addition to the lengths of derived routes (3rd argument
of \codett{route/4}), we can construct a proof tree
(4th argument).
An~atom \codett{prolog:A} leads to an embedded top--down
call of the goal \codett{A} in \prolog.
The goal \codett{(L is N+M)} computes the sum \codett{L}
of the path lengths \codett{N} and \codett{M} in \prolog.
A goal of the form \codett{pt(T, ...)} constructs
a proof tree \codett{T}.
\begin{quote}
\begin{lstlisting}{}
route(X, Y, L, T) :-
   street(X, Y, L, T1),
   prolog:pt(T, t(route(X, Y, L), e, T1)).
route(X, Y, L, T) :-
   street(X, Z, N, T1), route(Z, Y, M, T2),
   prolog:(L is N+M),
   prolog:pt(T,
      t(route(X, Y, L), r, T1, T2, (L is N+M))).

street('KT', 'Wue', 15, T) :-
   prolog:pt(T, t(street('KT', 'Wue', 15), f1)).
street('Wue', 'Mue', 280, T) :-
   prolog:pt(T, t(street('Wue', 'Mue', 280), f2)).
\end{lstlisting}
\end{quote}

  \newcommand{\DB}{P}
  \newcommand{\tdb}{{\cal T}_\DB}

\begin{figure}[htb]
   \mbox{} \hspace*{25mm}
   \includegraphics[width=0.6\columnwidth]{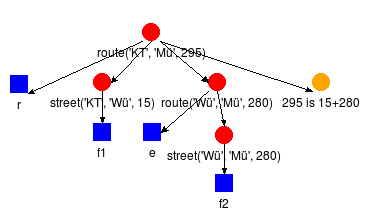}%
   \caption{Proof Tree.}
   \label{Proof Tree}
\end{figure}

In \ddbase, we have implemented a generalized $\tdb$--operator,
which can derive the following atom by a bottom--up evaluation.
It can be seen, that the evaluation derives a proof tree in
the last argument of the predicates \codett{route/4} and
\codett{street/4}.
This tree serves as an \emph{explanation} of the derived atoms,
a very useful concept known from expert systems.
E.g., this program will derive the atom shown in the following.
The last argument of the atom contains the proof tree,
which was automatically layouted and visualized in \ddbase,
see Figure~\ref{Proof Tree}.
\begin{quote}
\begin{lstlisting}{}
route(KT, Mue, 295,
   t(route(KT, Mue, 295), r,
      t(street(KT, Wue, 15), f1),
      t(route(Wue, Mue, 280), e,
         t(street(Wue, Mue, 280), f2))))
\end{lstlisting}
\end{quote}
\eat{
route(KT, Wue, 15,
   t(route(KT, Wue, 15), e,
      t(street(KT, Wue, 15), f1)))
route(Wue, Mue, 280,
   t(route(Wue, Mue, 280), e,
      t(street(Wue, Mue, 280), f2)))
street(KT, Wue, 15, t(street(KT, Wue, 15), f1))
street(Wue, Mue, 280, t(street(Wue, Mue, 280), f2))
}

\section{Ontologies with Rules in \swrl}

A hybrid information system can include ontologies of various
origins.
Before working with and -- and for designing such ontologies --
the knowledge engineer has to anaylse them and check them
for \emph{anomalies}.
In \ddbase, we use methods for detecting anomalies
in ontologies with rules, such as \textsc{Swrl} ontologies,
which we have investigated in~\cite{BaSei:10}.
For handling different versions of an ontology,
it is also possible to use well--known \emph{alignment} methods
to find out common parts and differences.

\subsection{Schema Graph for \xml}

\begin{figure}[h]
   \mbox{} \hspace*{7.5mm}
   \includegraphics[width=0.9\columnwidth]{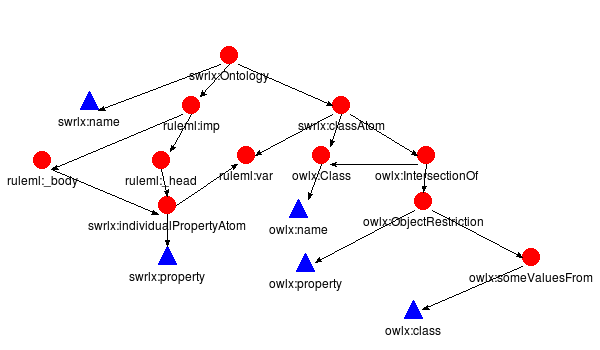}
   \caption{Schema Graph for \xml.}
   \label{Schema Graph}
\end{figure}

\swrl ontologies can be represented in \xml notation:
\begin{quote}
\begin{lstlisting}{}
<swrlx:Ontology swrlx:name="people">

<swrlx:classAtom> 
  <owlx:Class owlx:name="person"/>
  <ruleml:var>X</ruleml:var>
</swrlx:classAtom> 

<swrlx:classAtom> 
  <owlx:IntersectionOf>
    <owlx:Class owlx:name="person"/>
    <owlx:ObjectRestriction owlx:property="parent"> 
      <owlx:someValuesFrom owlx:class="Physician"/>
    </owlx:ObjectRestriction> 
  </owlx:IntersectionOf>
  <ruleml:var>Y</ruleml:var>
</swrlx:classAtom> 

<ruleml:imp> ... </ruleml:imp> 

</swrlx:Ontology>
\end{lstlisting}
\end{quote}
In \ddbase, a \swrl ontology can be visualized by the schema graph
of its \xml representation.
Since ontologies -- or \xml files in general -- can be very
large, it is helpful to get a short overview in advance,
see Figure~\ref{Schema Graph}.

\subsection{Rules in \textsc{Swrl}}

\swrl ontologies can have rules in \ruleml.
E.g., the following rule, which states that the brother of a parent
is an uncle, can be represented in \ruleml:
\begin{mquote}
   \Rule{uncle(X,Z)}{parent(X,Y) \wedge brother(Y,Z)}.
\end{mquote}
In the \xml representation, this could look like follows:
\begin{quote}
\begin{lstlisting}{}
<ruleml:imp> 
  <ruleml:_body> 
    <swrlx:individualPropertyAtom
      swrlx:property="parent"> 
      <ruleml:var>X</ruleml:var>
      <ruleml:var>Y</ruleml:var>
    </swrlx:individualPropertyAtom> 
    <swrlx:individualPropertyAtom
      swrlx:property="brother">
      ...
  </ruleml:_body> 
  <ruleml:_head> 
    <swrlx:individualPropertyAtom
      swrlx:property="uncle">
      ...
  </ruleml:_head> 
</ruleml:imp> 
\end{lstlisting}
\end{quote}

\subsection{Syntax and Semantics}

The syntax for \swrl in this section abstracts from any exchange syntax
for \owl and thus facilitates access to and evaluation of the language.
This syntax extends the abstract syntax of \owl described in the
\owl Semantics and Abstract Syntax document \cite{OWL S&AS}.
This abstract syntax is not particularly readable for rules.
Thus, examples will thus often be given in an informal syntax,
which will neither be given an exact syntax nor a mapping to any
of the fully--specified syntaxes for \swrl.

The \emph{abstract syntax} is specified here by means of a version
of Extended BNF, very similar to the EBNF notation used for \xml.
Terminals are quoted;
non-terminals are bold and not quoted. Alternatives are either separated by
vertical bars (|) or are given in different productions. Components that can
occur at most once are enclosed in square brackets; components that can
occur any number of times (including zero) are enclosed in curly braces.
Whitespace is ignored in the productions here.
Names in the abstract syntax are RDF URI references.
%
\eat{
These names may be abbreviated into qualified names.
using one of the following namespace names:
\begin{quote}
\begin{tabular}{|l|l|} \hline
Namespace name	& Namespace \\ \hline \hline
rdf	& http://www.w3.org/1999/02/22--rdf--syntax--ns\# \\ \hline
rdfs	& http://www.w3.org/2000/01/rdf--schema\# \\ \hline
xsd	& http://www.w3.org/2001/XMLSchema\# \\ \hline
owl	& http://www.w3.org/2002/07/owl\# \\ \hline
\end{tabular}
\end{quote}
}
The meaning of some constructs in the abstract syntax will be
informally described.
The formal meaning of these constructs can be defined
via an extension of the \owl DL model--theoretic semantics
\cite{OWL S&AS}.

An \owl ontology in the abstract syntax contains a sequence of axioms and facts.
Axioms may be of various kinds, e.g., subClass axioms and equivalentClass
axioms. It is proposed to extend this with rule axioms.
Similar to what is usual in logic programming,
a rule axiom consists of an antecedent (body) and a consequent (head),
each of which consists of a (possibly empty) set of atoms.
Antecedents and consequents
are treated as the conjunctions of their atoms.
\begin{quote} \tt
rule ::= 'Implies(' \{ annotation \} \\ \hspace*{10mm}
    antecedent consequent ')' \\
antecedent ::= 'Antecedent(' \{ atom \} ')' \\
consequent ::= 'Consequent(' \{ atom \} ')'
\end{quote}
Rules with an empty antecedent can be used to provide
unconditional facts; however such unconditional facts are better
stated in \owl itself, i.e., without the use of the rule construct.
Rules with conjunctive consequents could easily be
transformed -- via the Lloyd--Topor transformations
\cite{Lloyd:87} -- into multiple rules each with an atomic consequent.
\eat{
\begin{quote} \tt
atom ::= description '(' i--object ')' \\ \hspace*{5mm}
   | individualvaluedPropertyID \\ \hspace*{15mm}
      '(' i--object i--object ')' \\
   \hspace*{5mm}
   | datavaluedPropertyID '(' i--object d--object ')' \\ \hspace*{5mm}
   | same\_as '(' i--object i--object ')' \\ \hspace*{5mm}
   | different\_from '(' i--object i--object ')'
\end{quote}
}
Atoms can be of the form {\it c(X)}, {\it p(X,Y)}, {\it same\_as(X,Y)}
or {\it different\_from(X,Y)}, where {\it c} is an \owl description,
{\it p} is an \owl property, and {\it X,Y} are either variables,
\owl individuals or \owl data values.
In the context of \owl Lite, descriptions in atoms
of the form {\it c(X)} may be restricted to class names.
Informally,
\bi
   \item
an atom {\it c(X)} holds,
if {\it X} is an instance of the class description {\it c},
   \item
an atom {\it p(X,Y)} holds,
if {\it X} is related to {\it Y} by property {\it p},
   \item
an atom {\it same\_as(X,Y)} holds,
if {\it X} is interpreted as the same object as {\it Y}, and
   \item
an atom {\it different\_from(X,Y)} holds
if {\it X} and {\it Y} are interpreted as different objects.
\ei
The latter two forms can be seen as \emph{syntactic sugar}:
they are convenient, but do not increase the expressive power
of the language.
\eat{
\begin{quote} \tt
i--object ::= i--variable | individualID \\
d--object ::= d--variable | dataLiteral
\end{quote}
\begin{quote} \tt
i--variable ::= 'I--variable(' URIreference ')' \\
d--variable ::= 'D--variable(' URIreference ')'
\end{quote}
}
Atoms may refer to individuals, data literals, individual variables or data
variables.
As in \prolog or \datalog, variables are treated as universally
quantified, with their scope limited to a given rule.
Only variables occuring in the antecedent of a rule
may occur in the consequent (range--restrictedness).
This condition does not, in fact, restrict the expressive
power of the language,
because existentials can already be captured in \owl using
someValuesFrom restrictions.
\eat{
In abstract syntax, also restrictions,
such as $\leq_1(\mathit{process(X)})$ could be represented as
\begin{quote}
   restriction(process maxCardinality(1)).
\end{quote}
}


While the abstract EBNF syntax is consistent with the \owl specification,
and is useful for defining \XML and RDF serialisations, it is rather
verbose and not particularly easy to read.
Deductive databases therefore often use a relatively informal
\emph{human readable} form.
In this syntax, a rule has the form
   $\Rule{\alpha}{\beta}$,
where the antecedent~$\beta$ is a conjunction of atoms and
the consequent~$\alpha$ is a single atom.
In standard convention, variables are indicated by prefixing
them with a question mark;
in this paper, however, we represent them in \prolog convention
with strings starting with a capital letter.
Then, a rule asserting that the composition
of parent and brother properties implies the uncle property
would be written as
\begin{mquote}
   \Rule{uncle(X,Z)}{parent(X,Y) \wedge brother(Y,Z)}.
\end{mquote}
%
%
%
%
An even simpler rule would be to assert that students are persons:
\begin{mquote}
   \Rule{person(X)}{student(X)}.
\end{mquote}
However, this kind of use for rules in \owl just duplicates the \owl subclass
facility. It is logically equivalent to write instead
   {\it Class(Student partial Person)}
or
   {\it SubClassOf(Student Person)}
which would make the information directly available to an \owl reasoner.
A very common use for rules is to move property values from one individual to a
related individual.

\subsection{Hybrid Information Systems}

For collaborations across disciplines,
hybrid information systems using data and techniques
from many different sources with no preexisting agreement
about the semantics of the processes or data is important. 
The infrastructure must provide general purpose mechanisms
for annotating (i.e., making assertions about), discovering,
and reasoning about processes and data.
Some of the \emph{inferences} require additional reasoning
beyond that supported by \owl and \swrl.
Also graphs such as the \emph{provenance graph} are very
useful here for representing causal relationships.
The \emph{Open Provenance Model (OPM)} defines logical
constraints on the provenance graph~\cite{McGraFu:08}.
Some constraints cannot be expressed in \owl, but
can be expressed based on \swrl rules.

PROV is a specification that provides a vocabulary
to interchange provenance information.
It defines a core data model for the interchange of
\emph{provenance} on the web;
it allows for building representations of the entities,
people and processes involved in producing a piece
of data in the world. 
The provenance of digital objects represents their
origins;
the records of a PROV specification can describe
the entities and activities involved in producing
and delivering or otherwise influencing a given object.
Provenance can be used for many purposes,
such as understanding how data was collected
so it can be meaningfully used,
determining ownership and rights over an object,
making judgements about information to determine
whether to trust it,
verifying that the process and steps used
to obtain a result complies with given requirements,
and reproducing how something was generated.

The Open Provenance Model (OPM) provides a case study
for how to use Semantic Web technology and rules to
implement semantic metadata.
\cite{McGraFu:08} discusses a binding of the OPM written
in \owl with rules written in \swrl.
This allows for the development of hybrid
systems that use \owl, \swrl, and other semantic software,
interoperating through a shared space of RDF triples.
%
E.g., four $\mathit{derived}$ relations between two artifacts
(e.g., input and output datasets) can be inferred
from $\mathit{used}$ and $\mathit{generated\_by}$ to the same
process:
\begin{mquote}
   \mathit{derived\_sink(B, H)} \wedge
   \mathit{derived\_source(B, Y)} \wedge \\
   \mathit{derived\_account(B, D)} \wedge
   \mathit{derived\_account(B, G)} \Rulearrow \\ \hspace*{10mm}
      \mathit{artifact(Y)} \wedge
      \mathit{generated\_by\_artifact(C, Y)} \wedge \\ \hspace*{10mm}
      \mathit{process(X)} \wedge
      \mathit{generated\_by\_process(C, X)} \wedge \\ \hspace*{10mm}
      \mathit{account(D)} \wedge
      \mathit{generated\_by\_account(C, D)} \wedge \\ \hspace*{10mm}
      \mathit{relation(E)} \wedge
      \mathit{used\_process(E, X)} \wedge \\ \hspace*{10mm}
      \mathit{artifact(H)} \wedge
      \mathit{used\_artifact(E, H)} \wedge \\ \hspace*{10mm}
      \mathit{account(G)} \wedge
      \mathit{used\_account(E, G)} \wedge \\ \hspace*{10mm}
      \mathit{swrlx\!:\!create\_owl\_thing(B, X, C, E)}
\end{mquote}
\eat{
\begin{mquote}
   \Rule{style\_period(Z,Y)}{ \\ \hspace*{10mm}
      artist(X) \wedge artist\_style(X,Y) \wedge
      style(Y) \wedge creator(Z,X)}
\end{mquote}
It is useful to include \owl descriptions in rules, instead of using named
classes. The above rule could be augmented with a separate rule to provide
information about exclusivity of style (assuming that style is not always
exclusive).
\begin{mquote}
   \Rule{\leq_1(style\_period(Z))}{
      artist(X) \wedge \leq_1(artist\_style(X)) \wedge creator(Z,X)}
\end{mquote}
\begin{quote}
   restriction(style\_period maxCardinality(1)), \\
   restriction(artist\_style maxCardinality(1)).
\end{quote}
}
But several of the key constraints and inferences
of the OPM cannot be expressed in \owl and \swrl,
due to fundamental limits of the semantics of these languages.
E.g., it is not possible to modify the value of an asserted
property, or to write a rule discovering the number of times
an artifact is used, or to detect a cycle in the
provenance graph.
Storing the OPM records in triples makes it possible
to use other reasoning engines or languages
such as \prolog or \datalog to implement queries or inferences.

\owl and \swrl's \rdf representations provide a simple
and well--understood means of exchanging provenance
information with other tools, such as \rdf databases
and \emph{declarative} programming languages.
This hybrid system shows that Semantic Web technologies
are not only useful for provenance information but also
provide a base level of interoperability that can enable
loosely--coupled tools with varying levels of capability
and expressiveness.


\section{Querying Hybrid Knowledge Bases in \ddbase}

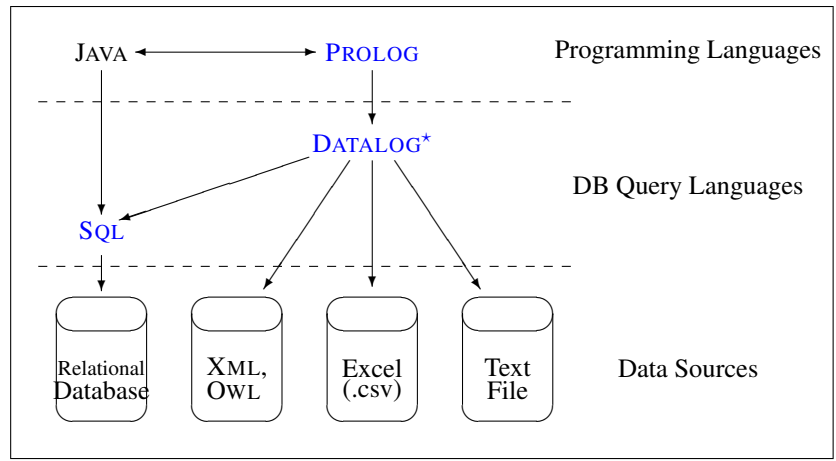
\begin{figure}[ht]
\begin{center} \small
\unitlength 6mm
\colorbox{white}{
\begin{picture}(17.5,9.7)(-0.67,-0.85)

\newsavebox{\dblpdatasource}
\savebox{\dblpdatasource}(0, 0)[c]{
   \put(0, 0){\oval(2, 0.75)[b]}
   \put(-1, 0){\line(0, 1){2}}
   \put(1, 0){\line(0, 1){2}}
   \put(0, 2){\oval(2, 0.75)}
}

\put(-1,-1){\framebox(18.2,10){}}

\put(5.6,5.67){\vector(-3,-1){4.2}}
\put(1,3.5){\vector(0,-1){0.8}}

\put(6.5,5.6){\vector(-2,-3){1.9}}
\put(7,5.6){\vector(0,-1){2.8}}
\put(7.5,5.6){\vector(2,-3){1.9}}

\put(7,7.6){\vector(0,-1){1.2}}
\put(1,7.6){\vector(0,-1){3.2}}
\put(1.75,8){\vector(1,0){4}}
\put(1.75,8){\vector(-1,0){0}}

\put(1,4){\makebox(0,0){\textcolor{blue}{\sql}}}
\put(7,6){\makebox(0,0){\textcolor{blue}{\datalogs}}}

\put(1,8){\makebox(0,0){{\java}}}
\put(7,8){\makebox(0,0){\textcolor{blue}{\prolog}}}

\put(14,1){\makebox(0,0){Data Sources}}
\put(14,5){\makebox(0,0){DB Query Languages}}
\put(14,8){\makebox(0,0){Programming Languages}}

\multiput(-0.4,3.25)(0.4,0){30}{\line(1,0){0.2}}
\multiput(-0.4,6.9)(0.4,0){30}{\line(1,0){0.2}}

\put(1,1){\usebox{\dblpdatasource}}
\put(1,1){\makebox(0,0){\scriptsize{Relational}}}
\put(1,0.5){\makebox(0,0){Database}}
\put(4,1){\usebox{\dblpdatasource}}
\put(4,1){\makebox(0,0){\xml,}}
\put(4,0.5){\makebox(0,0){\owl}}
\put(7,1){\usebox{\dblpdatasource}}
\put(7,1){\makebox(0,0){Excel}}
\put(7,0.5){\makebox(0,0){(.csv)}}
\put(10,1){\usebox{\dblpdatasource}}
\put(10,1){\makebox(0,0){Text}}
\put(10,0.5){\makebox(0,0){File}}

\end{picture}
}
\end{center}
   \caption{Hybrid Knowledge Bases in \ddbase.}
   \label{Hybrid Knowledge Bases in ddbase}
\end{figure}

Hybrid knowledge bases can be managed and queried
using logic programming techniques.
In the deductive database system \ddbase,
various representations of knowledge can be accessed,
see Figure~\ref{Hybrid Knowledge Bases in ddbase}.
The database query language \datalogs is an extension
of \datalog~\cite{Sei:09}, where logic programs in \prolog syntax
are evaluated bottom--up,
and embedded calls to \prolog are evaluated top--down.
\xml data from \xml databases or documents can be stored
in a term representation;
calls to \xml data based on path expressions
are evaluated in \prolog using the query, transformation and update
language \fnquery \cite{Sei:02}.

In \ddbase, we can compute joins of relational databases
and \xml documents in \prolog.
The following example is a modified version of the well--known
example from the book of Elmasri and Navathe \cite{ElNa:15}.
The atoms for \codett{employee/8} could, e.g., be derived using
\odbc from a relational database;
they could also stem from an ontology;
\swi \prolog \cite{Wielemaker:03b} offers a loader producing
\rdf triples, which can be transformed to \prolog facts easily.
\begin{quote} \tt\footnotesize
   \% employee(Name, SSN, Date, SEX, Salary, Super\_SSN, DNO) \\[3mm]
   employee('Borg',    11, date(1927,11,10), 'M', 55000, null, 1). \\
   employee('Wong',    22, date(1945,12,08), 'M', 40000, 11, 5). \\
   employee('Wallace', 33, date(1931,6,20), 'F', 43000, 11, 4). \\
   employee('Smith',   44, date(1955,1,09), 'M', 30000, 22, 5). \\
   ...
\end{quote}
Additionally, we work with the following \xml version of
the table \codett{works\_on/3}; a~row represents an
employee \codett{ESSN} working on a project \codett{PNO} a
number of \codett{HOURS}.
\begin{quote} \small
\begin{verbatim}
<table name="works_on">
   <row ESSN="11" PNO="20" HOURS="NULL"/>
   <row ESSN="22" PNO="2" HOURS="10.0"/>
   ...
</table>
\end{verbatim}
\end{quote}

The following query joins the atoms for \codett{employee/8}
with the rows in the \xml document \codett{works\_on.xml} in \ddbase.
The attribute value {\tt H} of the attribute {\tt 'HOURS'} of {\tt Row}
is an atom that has to be converted to a number {\tt HOURS}.
Clearly, this predicate fails for \codett{HOURS = null},
which is desired to ignore null values in aggregations in~\sql.
The handling of path expressions applied to
\xml documents has been described in \cite{Sei:02}.
The template {\tt [DNO, sum(HOURS)]} leads to a grouping on the
department numbers.
For every {\tt DNO}, the list {\tt Xs} of all corresponding {\tt HOURS}
is computed, and the sum by {\tt sum(Xs, Sum)};
thus, we obtain a standard result tuple {\tt [DNO, Sum]}.
\begin{quote}
\begin{verbatim}
?- ddbase_aggregate( [DNO, sum(HOURS)],
      ( employee(_, SSN, _,_,_,_,_, DNO),
        Row := doc('works_on.xml')/row::[@'ESSN'=SSN],
        H := Row@'HOURS', atom_number(H, HOURS) ),
      Tuples ).
\end{verbatim}
\end{quote}
\begin{quote}
\begin{verbatim}
Tuples = [[1, 0.0], [4, 115.5], [5, 140.0]]
\end{verbatim}
\end{quote}
A query optimizer of \ddbase could rearrange the Goal in the
second argument of the \prolog atom for
\verb+ddbase_aggregate/3+ by changing the order of the calls
to the predicate \verb+employee/8+ provided by \odbc and
the \xml document \verb+works_on.xml+.
It might be the best to first completely load the table
\textsc{Employee} from the relational database to \prolog
using \odbc
and to index it on the second argument position, which holds
the~{\tt SSN}.
Then, in a single pass through the \xml document, the working hours
can be obtained using a path expression in \fnquery from the \xml rows,
and the corresponding department numbers of the employees
can be obtained using the index.

\section{Final Remarks}

We have developed a \prolog--based deductive database system
\ddbase for hybrid knowledge bases.
Knowledge sources with different types of
knowledge representation~-- including relational databases, \xml,
\swrl knowledge bases~-- can be managed in \ddbase.

The transition from relational databases to deductive databases
brings in recursive rules, and thus generalizes the concept
of  views.
\swrl ontologies add further ideas from artificial
intelligence;
ontologies can be augmented by rules
to enhance expressiveness.
PROV uses \swrl to model the provenance of digital data.

In \ddbase, we are building a system for handling hybrid queries
in a deductive data\-base.
A query optimizer should extend relational systems,
and it should be able to handle relational data and rules
together with \xml data and ontologies.

\bibliographystyle{elsart-num}
\bibliography{paper}

\end{document}